# Numerical study of incommensurate and decoupled phases of spin-1/2 chains with isotropic exchange $J_1$, $J_2$ between first and second neighbors

Zoltán G. Soos,[1] Aslam Parvej[2] and Manoranjan Kumar[2]

[1]Department of Chemistry, Princeton University, Princeton, New Jersey 08544, USA

[2]S.N. Bose National Centre for Basic Sciences. Block-JD, Sector-III, Kolkata 700098, India

**Abstract**

The spin-1/2 chain with isotropic exchange $J_1$, $J_2 > 0$ between first and second neighbors is frustrated for either sign of $J_1$ and has a singlet ground state (GS) for $J_1/J_2 \geq -4$. Its rich quantum phase diagram supports gapless, gapped, commensurate (C), incommensurate (IC) and other phases. Critical points $J_1/J_2$ are evaluated using exact diagonalization (ED) and density matrix renormalization group (DMRG) calculations. The wave vector $q_G$ of spin correlations is related to GS degeneracy and obtained as the peak of the spin structure factor $S(q)$. Variable $q_G$ indicates IC phases in two $J_1/J_2$ intervals, [– 4, –1.24] and [0.44, 2], and a C-IC point at $J_1/J_2 = 2$. The decoupled C phase in [–1.24, 0.44] has constant $q_G = \pi/2$, nondegenerate GS, and a lowest triplet state with broken spin density on sublattices of odd and even numbered sites. The lowest triplet and singlet excitations, $E_m$ and $E_\sigma$, are degenerate in finite systems at specific frustration $J_1/J_2$. Level crossing extrapolates in the thermodynamic limit to the same critical points as $q_G$. The $S(q)$ peak diverges at $q_G = \pi$ in the gapless phase with $J_1/J_2 > 4.148$ and quasi-long-range order (QLRO($\pi$)). $S(q)$ diverges at $\pm \pi/2$ in the decoupled phase with QLRO($\pi/2$), but is finite in gapped phases with finite-range correlations. Numerical results and field theory agree at small $J_2/J_1$ but disagree for the decoupled phase with weak exchange $J_1$ between sublattices. Two related models are summarized: one has an exact gapless decoupled phase with QLRO($\pi/2$) and no IC phases; the other has a single IC phase without a decoupled phase in between.

## 1. Introduction

The $J_1$-$J_2$ model with isotropic exchange $J_1$, $J_2$ between first and second neighbors is the prototypical frustrated spin-1/2 chain with a dimer phase, also called a bond-order-wave phase.[1-16] The Hamiltonian with periodic boundary conditions (PBC) is

$$H(J_1, J_2) = J_1 \sum_r \vec{s}_r \cdot \vec{s}_{r+1} + J_2 \sum_r \vec{s}_r \cdot \vec{s}_{r+2} \qquad (1)$$

There is one spin per unit cell and the total spin S is conserved. The limit $J_2 = 0$, $J_1 > 0$ is the linear Heisenberg antiferromagnet (HAF) with nondegenerate ground state and quasi-long-range order (QLRO) at wave vector $q = \pi$. The limit $J_1 = 0$, $J_2 > 0$ corresponds to HAFs on sublattices of odd and even numbered sites, QLRO at $q = \pi/2$ and frustration for either sign of $J_1$. The parameter $g = J_2/J_1$ quantifies the competition between first and second neighbor exchange. Sandvik[17] has reviewed numerical studies of the HAF and related spin chains, including $H(J_1,J_2)$ at $g < 1$. An earlier review by Lecheminant[18] addresses frustrated 1D spin systems mainly in terms of field theory, also for $g < 1$.

In addition to extensive HAF results, the exact ground state (GS) is known at $g_{MG} = 1/2$, the Majumdar-Ghosh[1] point, and at $J_1/J_2 = -4$, $J_2 > 0$, the quantum critical point P1 discussed by Hamada et al.[7] and shown in Fig. 1. Okamoto and Nomura[9] used exact diagonalization (ED) of finite systems, level crossing and extrapolation to find the critical point P4 at $1/g_{ON} = 4.148$ in Fig. 1. The dimer phase has doubly degenerate GS, broken inversion symmetry at sites and finite energy gap $E_m$ to the lowest triplet state. The dimer phase and P4 were the initial focus of theoretical and numerical studies.[1-18] Attention has recently shifted to the $J_1$-$J_2$ model with ferromagnetic $J_1 < 0$ that is the starting point for the magnetic properties of Cu(II) chains in some cupric oxides.[19-22] Moreover, models in an applied magnetic field or with anisotropic exchange have multipolar, vector chiral and exotic phases.[23-28] Furukawa et al.[28] discuss both anisotropic and isotropic exchange using field theory and numerical methods and note that less is known about the $J_1 < 0$ sector.

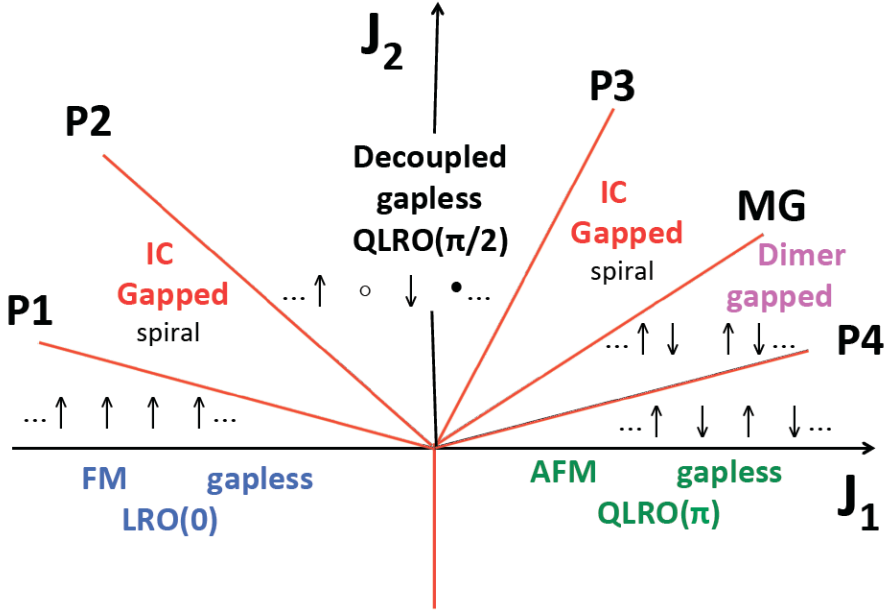

Fig. 1. Quantum phase diagram of $H(J_1,J_2)$, Eq. 1. The $J_1/J_2$ values at the critical points are P1 = – 4, P2 = –1.24, P3 = 0.44 and P4 = 4.148. The exact point P1 is between a gapless FM phase and a gapped incommensurate (IC) phase. The gapless decoupled phase is between P2 and P3; open and closed circle denote spins pointing in and out of the plane. The gapped IC phase extends to the MG point, $J_1 = 2J_2$, and the dimer phase to P4 = 4.148, beyond which lies a gapless AFM phase.

The critical points P2 and P3 in Fig. 1 are obtained in the present study. An earlier estimate (P2 = –1.2, P3 = 0.45) was based[29] on level crossing and divergent structure-factor peaks. We consider the phase diagram of the $J_1$-$J_2$ model with particular attention to P2 and P3, the incommensurate or spiral phases with spin correlations of finite range, the decoupled phase with QLRO($\pi/2$), and the commensurate to incommensurate (C-IC) point[30,31] at $J_1/J_2 = 2$. We present numerical evidence such as the degeneracy of the GS to identify IC phases, the periodicity of spin correlations and the spin densities of the triplet state using ED for systems up to N = 28 spins and density matrix renormalization group (DMRG) calculations.[32,33] DMRG returns accurate GS energies and spin correlation functions C(r) in systems of a few hundred spins. The combination of ED and DMRG in finite systems affords a detailed picture of IC and decoupled phases. The thermodynamic limit is deferred as long as possible.

Field theories introduce continuous operators for spin chains at the beginning and follow renormalization group (RG) flows to distinguish between gapped and gapless phases. Exponentially small energy gaps or long-range spin correlations are beyond the reach of approximate numerical methods, and that is the case when $J_1/J_2$ is small or negative. Direct comparison is limited to systems with short-range spin correlations. Other comparisons are needed to support field theory or to assess conflicting results for IC and dimer phases. However, all field theories[12-14,18,23-26,28] find that the QLRO($\pi/2$) phase in Fig. 1 is limited to the point $J_1 = 0$. We question the assertion that arbitrarily small $J_1$ suppresses the QLRO($\pi/2$) phase while finite $J_2/J_1 = 0.2411$ is needed to suppress the QLRO($\pi$) phase. Weak exchange $J_1$ between HAFs on sublattices poses interesting and unresolved challenges, akin to dispersion forces, that merit closer attention.

There are basic differences at arbitrarily small gaps. Gapless critical phases at $J_1 = 0$ or $J_2 = 0$ have nondegenerate GS and divergent structure factor peaks. Gapped phases have doubly degenerate GS and finite structure factor peaks. We show that variable q in IC phases provides direct information about GS degeneracy and that the lowest triplet in the QLRO($\pi/2$) phase has broken sublattice spin densities. The physical picture in Fig. 1 extends the QLRO($\pi/2$) phase of HAFs on sublattices to small frustration $J_1$ of either sign, just as the QLRO($\pi$) phase is stable against small frustration $J_2$. Increasing $J_1/J_2 > 0$ induces a quantum transition at P3 to a gapped IC phase with $q \geq \pi/2$ that terminates at the MG point, $J_1/J_2 = 2$. Decreasing $J_1/J_2 < 0$ gives a transition at P2 to an IC phase with $q \leq \pi/2$ that terminates at $J_1/J_2 = -4$. We use other numerical results than exponentially small energy gaps to obtain the quantum phase diagram in Fig. 1.

The paper is organized as follows. GS degeneracy and inversion symmetry are related in Section 2 to the wave vector $q_G$ of ground state correlations. The structure factor $S(q)$, the Fourier transform of GS spin correlations, of finite systems is a discrete function. Its peak $S(q^*)$ is shown in Section 3 to occur at $q^* = q_G$ except near the C-IC point. DMRG yields $q_G$ in systems of several hundred spins using spin correlations instead of energy degeneracy. Level crossing of excited states is combined with $q_G$ in Section 4 and extrapolated to the thermodynamic limit to estimate the critical points P2 and P3. The decoupled phase is characterized in Section 5 using the spin densities and sublattice spin of the lowest triplet state. The magnitude of $S(q)$ peaks at $\pi$ or $\pi/2$ with increasing system size are compared in Section 6. The GS expectation value $\langle S_A^2 \rangle = \langle S_B^2 \rangle$ of the square of sublattice spin is related in Section 7 to the asymmetry of P2 and P3 about $J_1 = 0$. Spin correlations $C(2r)$ within sublattices and $C(2r-1)$ between sublattices are compared. Two extensions of the $J_1$-$J_2$ model are sketched in Section 8. The first is an analytical model in which IC phases are suppressed and the decoupled phase expands to $J_1/J_2 = \pm 4\ln 2$. The second has frustrated sublattices and a single IC phase without an intervening QLRO($\pi/2$) phase. In the Discussion we summarize the numerical evidence for the phase diagram in Fig. 1 and compare the field theoretical expression for $C(r)$ in gapped IC phases to DMRG results.

**2. Ground state spin correlations**

An even number of spins is assumed in spin-1/2 chains with isotropic exchange to ensure integer $S \leq N/2$. We take $N = 4n$ in order to have integer spin $S_A, S_B \leq n$ on sublattices of odd and even numbered sites. The $J_1 = 0$ limit of Eq. 1 is then 2n-spin HAFs, which is quite different from half-integer $S_A, S_B$ when $N/2$ is an odd integer. The spontaneously broken symmetry is inversion $\sigma$ at sites. Finite systems of 4n spins with PBC also have inversion symmetry $\sigma'$ at the center of bonds. Open boundary conditions break $\sigma$ symmetry and limit $\sigma'$ to the central bond. While this does not matter in the thermodynamic limit, the issue would not come up if there were accurate results in that limit. The size limitations of ED are partly compensated by access to excited states. DMRG yields accurate GS properties[32,33] of much larger systems. We use DMRG with PBC and four spins added per step.[29,34]

The GS is nondegenerate in $J_1$-$J_2$ models of 4n spins with PBC except at 2n values[15,27] of $J_1/J_2$. The wave vector is k = 0 or π in the σ = 1 or –1 sectors. We focus below on the wave vector $q_G$ of GS spin correlations instead of k. As sketched in Fig. 1, spin correlations in the ferromagnetic GS with $J_1/J_2 \leq -4$ have long-range order (LRO) at $q_G = 0$. On the antiferromagnetic side, the GS has short-range spin correlations at $q_G = \pi$ for $J_1/J_2 \geq 2$ and QLRO at $q_G = \pi$ for $J_1/J_2 > 4.148$. The point $J_1 = 0$ with HAFs on sublattices has QLRO at $q_G = \pm \pi/2$. The GS of *classical* spins in Eq. 1 is a spiral phase with LRO($q_{cl}$) in the interval $-4 \leq J_1/J_2 \leq 4$. The pitch angle $q_{cl}$ between adjacent spin is given by $4\cos q_{cl} = -J_1/J_2$ and ranges from $q_{cl} = 0$ to $\pm \pi$ with increasing $J_1/J_2$. The same range of $q_G$ occurs for *quantum* spins in which fluctuations suppress LRO and there is no simple relation to $J_1/J_2$. The evolution of $q_G$ with increasing $J_1/J_2$ holds more generally for spin-1/2 chains with isotropic exchange and $q_G = 0$ or π at large negative or positive $J_1$.

Spin correlations $\langle s_0 \cdot s_r \rangle$ are GS expectation values that are commensurate in finite systems and limited to r ≤ 2n for N = 4n sites. PBC leads to discrete q in the first Brillouin zone

$$q = \frac{\pi r}{2n}, \qquad r = 0, \pm 1, ....., 2n \qquad (2)$$

The periodicity of GS correlations in the $J_1$-$J_2$ model increases from $q_G = 0$ to π, or decreases from $q_G = 0$ to –π, in 2n steps of π/2n between P1 at $J_1/J_2 = -4$ and the MG point at $J_1/J_2 = 2$. The end points are exact and hold in the thermodynamic limit.

The GS degeneracy at $J_1/J_2 = 2$ is between two singlets, the Kekulé valence bond (VB) diagrams for singlet-pairing of adjacent spins,

$$\begin{aligned} |K1\rangle &= (1,2)(3,4)....(4n-1,4n) \\ |K2\rangle &= (2,3)(4,5)....(4n,1) \end{aligned} \qquad (3)$$

where $(1,2) = (\alpha_1\beta_2 - \beta_1\alpha_2)/\sqrt{2}$ at sites 1 and 2. The diagrams are related by inversion at sites. The inversion symmetry is also broken at $J_1/J_2 = -4$ and altogether at $2n$ points $g_p(4n) = J_2/J_1$, $p = 1,2,\ldots 2n$.[15,27] The $g_p$ are known up to $N = 28$ and are accessible[17] for $N = 32$. The Kekulé diagrams are asymptotically orthogonal, with overlap $\langle K1|K2\rangle = -2^{-2n+1}$ given by Pauling's island counting rule.[35] Spin correlations in small systems at $J_1/J_2 = 2$ depend on overlap,[36] but overlap is negligible for $N > 24$ and $\langle s_0 \cdot s_r\rangle = 0$ for $r \geq 2$ at the MG point.

The ferromagnetic GS for $J_1/J_2 \leq -4$ and singlet GS for $J_1/J_2 \geq 2$ are in the $\sigma = 1$ sector for arbitrarily large $N = 4n$. The GS transforms as $\sigma = -1$ at $q_G = \pi r/2n$ when $r$ is an odd integer and as $\sigma = 1$ when $r$ is an even integer. Figure 2 shows $q_G$ for $N = 24$ as a function of $J_1/J_2$ with 12 steps at $g_p(4n)$. The staircase goes from $q_G = 0$ to $\pi$. An equivalent staircase runs from $q_G = 0$ to $-\pi$ $(= \pi)$. The stairs have equal risers $\pi/2n$ and variable steps governed by $g_p(4n)$, both of which depend on system size. The onset and termination of IC phases at $J_1/J_2 = -4$ and 2 are exact. The long step or plateau at $q_G = \pi/2$ in Fig. 2 is between $g_n(4n)$ with $J_1 < 0$ and $g_{n+1}(4n)$ with $J_1 > 0$. Another special feature of this step becomes apparent on doubling the system from N to 2N, or from 24 to 48 in Fig. 2. The wave vectors $\{q\}_N$ also appear in $\{q\}_{2N}$ of the larger system that has N additional $q$'s. The additional wave vectors are midway between every $\{q\}_N$ except at $q = \pi/2$ where there are two new steps at $q = \pi/2 \pm \pi/N$.

GS degeneracy and inversion symmetry specify the periodicity $2\pi/q_G$ of spin correlations in finite $J_1$-$J_2$ models. The wave vectors are uniformly distributed while the degeneracies are not. The degeneracy density between $0 \leq J_1/J_2 \leq 2$ is twice that between $-4 \leq J_1/J_2 \leq 0$. More important is the possibility of an interval without degeneracy. What is the fate of the $q_G = \pi/2$ plateau in the thermodynamic limit? A plateau implies a C phase, as in Fig. 1, between two IC phases. Small deviations from $q_G = \pm \pi/2$, on the other hand, would indicate a C point at $J_1 = 0$ between IC phases. Exact degeneracies require ED and are therefore limited to small systems. To follow the evolution of $q_G$ with increasing $J_1/J_2$, we use DMRG and GS spin correlations instead of energy degeneracy.

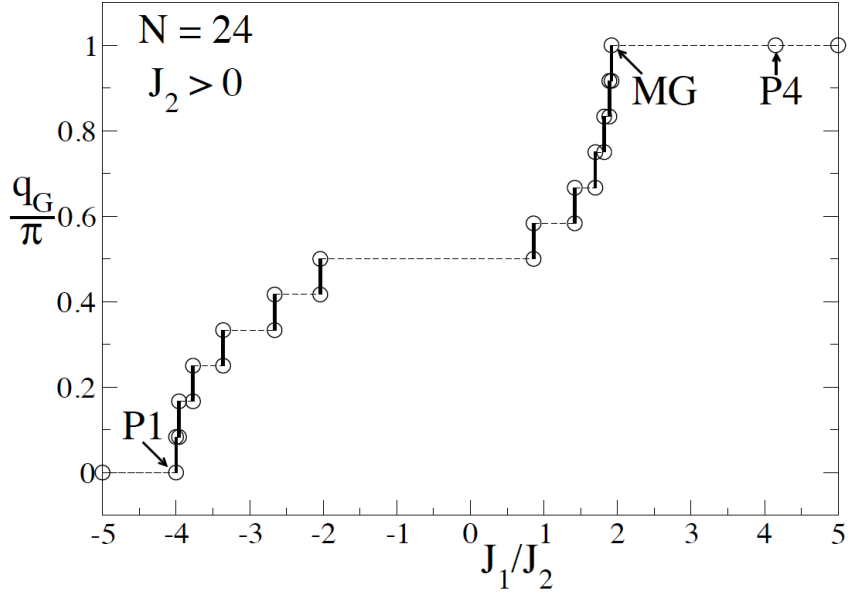

Fig. 2. ED results for the wave vector $q_G$ of GS correlations of the $J_1$-$J_2$ model with 24 spins in Eq. 1. The singlet GS is degenerate at 12 points in sectors that are even and odd under inversion at sites; $q_G$ switches between $\sigma = \pm 1$ with increasing $J_1/J_2$ starting with $q_G = 0$ at $J_1/J_2 = -4$ and ending with $q_G = \pi$ at $J_1/J_2 = 2$, both exact in the thermodynamic limit. The critical point P4 = 4.148 is to the gapless phase that includes the linear Heisenberg antiferromagnet at $J_2 = 0$.

## 3. Spin structure factor

The static structure factor $S(q)$ is the Fourier transform of spin correlations in the ground state

$$S(q) = \sum_r \langle \vec{s}_0 \cdot \vec{s}_r \rangle \exp(iqr) \qquad (4)$$

The wave vectors in Eq. 2 are for 4n spins and one spin per unit cell. $S(q)$ is even in q and symmetric about $q = 0$ and $\pi$. $S(0)$ is the sum of $\langle \vec{s}_0 \cdot \vec{s}_r \rangle$ over r, and $S(0) = 0$ when the GS is a singlet. The discrete function S(q) has peaks at ± q* over a finite $J_1/J_2$ interval. Bursill et al.[11] discussed an effective periodicity 2π/q* based on S(q) peaks. Except close to $J_1/J_2$ = 2, IC phases have $q_G$ = q*. Instead of energy degeneracy and symmetry, however, q* is based on GS spin correlations and is much more convenient to evaluate. First, we are seeking $q_G$ as a function of $J_1/J_2$ in the thermodynamic limit rather than staircases such as Fig. 2 in large systems. Second and more importantly, DMRG is an excellent GS approximation that yields spin correlations in large systems.

Broken inversion symmetry is the motivation for evaluating average spin correlations as

$$C(r) = \left(\langle \vec{s}_0 \cdot \vec{s}_r \rangle + \langle \vec{s}_0 \cdot \vec{s}_{-r} \rangle\right)/2 \qquad (5)$$

The two expectation values are equal by translational symmetry for nondegenerate GS, for example in the σ = 1 or –1 sectors, but are not equal for arbitrary linear combinations of degenerate GS. Indeed, the order parameter of the dimer phase in Fig. 1 is the difference between the two expectation values with r = 1 in Eq. 5 and can be evaluated directly in finite $J_1$-$J_2$ models at points where the GS is doubly degenerate.[15,27] The Kekulé diagrams have $\langle \mathbf{s}_0 \mathbf{s}_1 \rangle$ = –3/4 and 0 at nearest neighbors. The average C(r) enters in S(q) since the sum in Eq. 4 is over sites both to the right and left. Inversion symmetry is not specified in our DMRG algorithm. We compute both expectation values in Eq. 5 and take the average or difference as required by the problem being addressed.

The nodal structure of C(r) for $J_1$ < 0 confirms that $q_G$ = q*. The S(q) peak is at the Fourier component that matches the sign changes of C(r). Peaks at q* = ± π/2n close to $q_G$ = 0 at P1 are well resolved for any system size since S(0) = 0 in the singlet GS with σ = –1 and two sign changes of C(r). The peaks jumps to q* = ± π/n at $g_2(4n)$ when the GS is even under inversion and C(r) changes sign four times. We verified for N = 24 that q* follows $q_G$ exactly when $J_1$ < 0 and C(r) changes sign up to 2n times. We have C(2r–1) =

0 at $J_1 = 0$ and $C(2r) \propto (-1)^r$. The $q^* = \pi/2$ peak is reached well before $J_1 = 0$, however, at small rather than vanishing correlations between spins in different sublattices.

We also find $q^* = q_G$ for $J_1 > 0$ except near the MG point, $J_1/J_2 = 2$, where the exact structure factor is

$$S_{MG}(q) = 3(1-\cos q)/4 \qquad (6)$$

The size dependence is entirely in the discrete values of q, with $q^* = \pi$ and $S_{MG}(\pi) = 3/2$. A broad profile in reciprocal space is readily understood as extremely short range C(r) in real space. The profile narrows for $J_1/J_2 > 2$ as the range of correlations increases;[29] $S(\pi)$ increases and diverges in the QLRO($\pi$) phase with $J_1/J_2 \geq 4.148$.

On the other hand, the GS in the $\sigma = -1$ sector at the MG point has $q_G = \pi - \pi/2n$ and is the C-IC point of the $J_1$-$J_2$ model, as noted[30,31] previously for spin chains with an exact point. Finite $S(\pi)$ leads to overlapping profiles at $q_G = \pi \pm r\pi/2n$ that are first resolved in the thermodynamic limit at the Lifshitz point where $S''(\pi) = 0$. Bursill et al.[11] found $(J_2/J_1)_L = 0.52063(6)$ using DMRG with open boundary conditions. $S''(\pi) = 0$ requires that the spin correlations at $(J_1/J_2)_L$ satisfy

$$0 = -4n^2 C(2n) - 2\sum_{r=1}^{2n-1} r^2 (-1)^r C(r) \qquad (7)$$

Short range correlations make it possible to evaluate $(J_1/J_2)_L$ accurately. We find $(J_2/J_1)_L = 0.52066(2)$ for $N > 50$ in excellent agreement with the earlier result.[11] The resolved peaks at $\pm q^*$ separate for $J_1/J_2 < 1.92$ and merge with $\pm q_G$. We have $q^* = q_G$ for $q_G \leq 2\pi/3$ at $N = 24$ and expect similar merging of $q^*$ to $q_G$ in large systems since as shown in Section 7, $S(\pi)$ decreases rapidly with decreasing $J_1/J_2 < 2$.

DMRG and S(q) peaks yield $q_G$ in large systems. The evolution of $q^* = q_G$ with $J_1/J_2$ in IC phases is shown in Fig. 3 for $4n = 48$, 96 and 144. The point at $J_1 = 0$ is exact, as are $q_G = 0$ and $\pi$ for $J_1/J_2 \leq -4$ and $\geq 2$, respectively. The size dependence is confined to the $q_G = \pi/2$ plateau that defines an interval with nondegenerate GS in finite systems. The plateau is reached at $(J_1/J_2)_n$ in a step of $\pi/2n$ for $J_1 < 0$ and is left at $(J_1/J_2)_{n+1}$ in a step of $\pi/2n$ for $J_1 > 0$. These points are listed in Table 1 up to $N = 192$ and are shown in the insets of Fig. 3. The lines are linear extrapolations to the thermodynamic limit. The intercepts $J_1/J_2 = -1.24$ and $0.44$ are the $q_G$ estimates for P2 and P3 in Fig. 1. To check the accuracy, we recall that the sum $\Sigma_r C(r) = S(0)$ is zero in singlet states; it is $< 6 \times 10^{-4}$ up to $N = 144$ in Table 1 and about $3 \times 10^{-3}$ at $N = 192$. The linear behavior of $(J_1/J_2)_n$ and $(J_1/J_2)_{n+1}$ for $N > 48$ are directly related to system size since $q_G$ steps are known to be $\pm \pi/2n$.

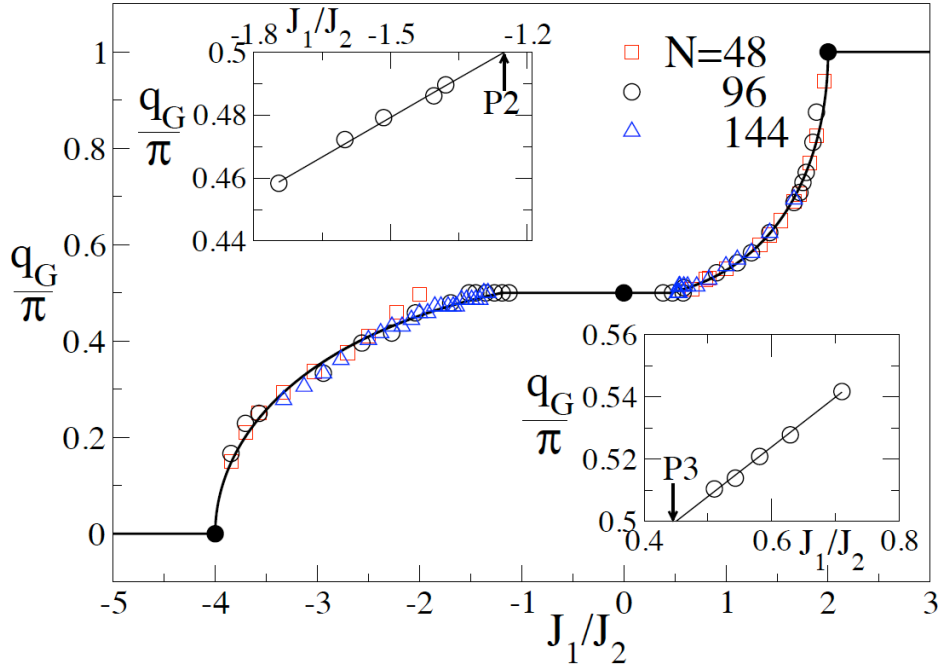

Fig. 3. DMRG results for the wave vector $q_G$ of GS correlations as a function of $J_1/J_2$ in models of N spins in Eq. 1. Closed circles are exact in the thermodynamic limit. The insets show the $J_1/J_2$ values in Table 1 at which the $q_G/\pi = 1/2$ plateau is reached and left in a step $2/N$ for $J_1 < 0$ and $J_1 > 0$; linear extrapolation gives the quantum critical points

P2 = –1.24 and P3 = 0.44. The intervals [– 4, P2] and [P3, 2] with variable $q_G$ are IC phases with degenerate GS. The C phase between P2 and P3 has a nondegenerate GS.

Table 1. DMRG results for the points $(J_1/J_2)_n$ and $(J_1/J_2)_{n+1}$ at which the $q_G = \pi/2$ plateau is reached and left in $J_1$-$J_2$ models of 4n spins.

| N = 4n | $(J_1/J_2)_n$ | $(J_1/J_2)_{n+1}$ |
|---|---|---|
| 24 | –2.033 | 0.868 |
| 48 | –1.745 | 0.710 |
| 72 | –1.600 | 0.629 |
| 96 | –1.515 | 0.581 |
| 144 | –1.439 | 0.543 |
| 192 | – 1.379 | 0.510 |
| ∞ | – 1.24 | 0.44 |

To conclude the analysis of $q_G$ in IC phases, we construct the solid line in Fig. 3. Classical spins account for the square root behavior at the exact point $J_1/J_2 = – 4$. Nomura and Murashima[33] suggested on general field theoretical grounds that $q \propto (g - g_c)^{1/2}$ near the C-IC point, with $J_2/J_1 \geq g_c = 1/2$. We expand instead about small $J_1/J_2$ and recall that the GS correlations of Eq. 1 have $q_G = 0, \pi/2$ and $\pi$, respectively, at $J_1/J_2 \leq – 4, 0$ and $\geq 2$. The combination of a square-root singularity at exact points and the $J_1 = 0$ constraint of $q_G = \pi/2$ suggests the expressions

$$\begin{aligned}
\frac{q_G}{\pi} &= A\left(4+\frac{J_1}{J_2}\right)^{1/2} \exp\left(-a\left(4+\frac{J_1}{J_2}\right)\right) \leq \frac{1}{2} & -4 \leq \frac{J_1}{J_2} \leq P2 \\
\frac{q_G}{\pi} &= \frac{1}{2} & P2 \leq \frac{J_1}{J_2} \leq P3 \quad (8) \\
\frac{q_G}{\pi} &= 1-B\left(2-\frac{J_1}{J_2}\right)^{1/2} \exp\left(-b\left(2-\frac{J_1}{J_2}\right)\right) \geq \frac{1}{2} & P3 \leq \frac{J_1}{J_2} \leq 2
\end{aligned}$$

The corresponding curve from $q_G = 0$ to $-\pi$ ($= \pi$) is $-q_G/\pi$. The parameters in Fig. 3 are A = 0.38, B = 0.565, a = 0.086 and b = 0.22. A and B refer to singularities that are known in the thermodynamic limit; a and b are negligible near the exact points but they matter for the critical points P2 and P3 that delimit the $q_G = \pi/2$ phase. Eq. 8 combines singularities at exact points and with the translational symmetry of the $J_1$-$J_2$ model before making a continuum approximation.

## 4. Level crossing

The GS degeneracy of finite $J_1$-$J_2$ models is between singlets with opposite inversion symmetry. We define $E_\circ$ and $E_m$ as the excitation energy to the lowest singlet and triplet, respectively. Both have finite-size contributions. Motivated by field theory, Okamoto and Nomura[9] argued that the gapped dimer phase with doubly degenerate GS must have two singlets below the lowest triplet. In finite systems, the singlet and triplet cross at $g^*(N)$ where $E_\circ = E_m$. They found[9] $g^*(N)$ exactly to N = 24, noted the weak size dependence and extrapolated to $g_{ON} = 0.2411 = 1/P4$.

The excitations $E_m$ and $E_\circ$ are well known at $J_2 = 0$. To lowest order in logarithmic corrections, Woynarovich and Eckle report[37]

$$E_m(N) = \frac{\pi^2}{2N}\left(1 - \frac{1}{2\ln N}\right) \tag{9}$$

Faddeev and Takhtajan show[38] that the triplet ($E_m$) and singlet ($E_\circ$) are degenerate in the infinite chain; they are the S = 1 and 0 linear combinations of two S = 1/2 kinks with identical dispersion relations. Combining $E_m(N)$ with coupling constants reported by Affleck et al.,[8] the difference $E_\circ(N) - E_m(N)$ is of order $1/(N\ln N)$, even smaller than $1/N$.

The $J_1 = 0$ limit of Eq. 1 has HAFs on sublattices. Now $E_m$ and $E_σ$ transform[29] with wave vector $k = \pm π/2$ and remain doubly degenerate for small $J_1/J_2$. The 9-fold degeneracy at $2E_m$ for a triplet on each sublattice corresponds to a singlet, a triplet and a quintet. The degeneracy is lifted for $J_1 > 0$ when only total spin is conserved. The singlet at excitation energy $^1E_{TT}$ has allowed crossings[29] with $E_σ$ and $E_m$ at finite $J_1/J_2$ that are shown in Fig. 4 up to $N = 28$. The HAF excitations above rationalize why both level crossings in finite systems extrapolate to $J_1/J_2 \sim 0.45$. The crossings $E_σ(N) = E_m(N)$ for $J_1 < 0$ are also shown to $N = 28$.

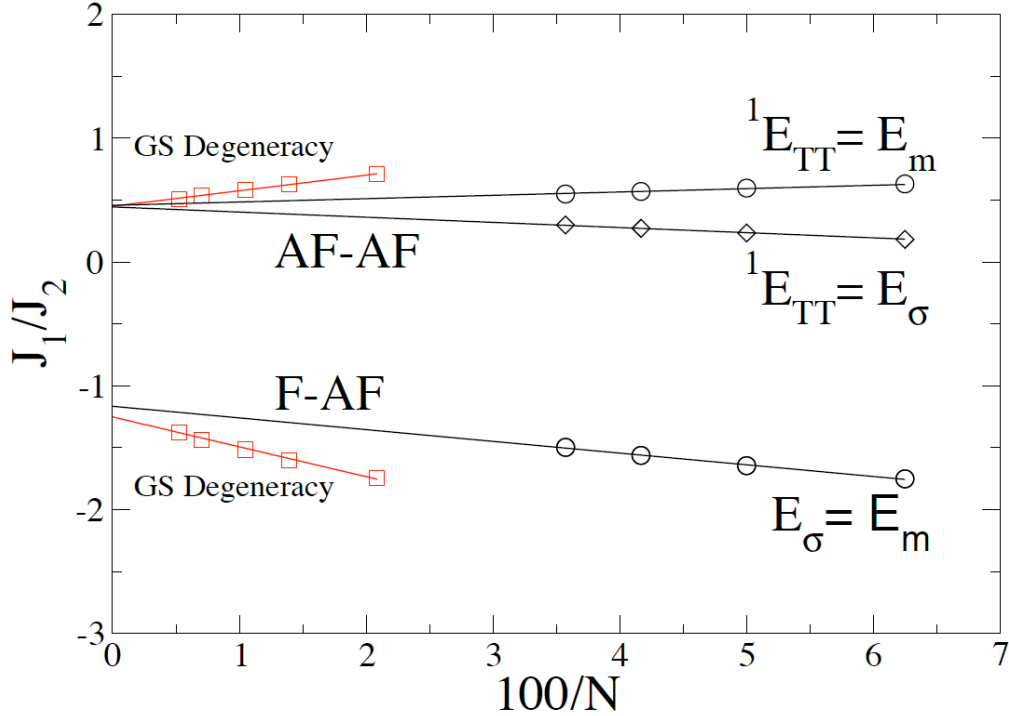

Fig. 4: ED results for level crossing and DMRG results for GS degeneracy in $J_1$-$J_2$ models with $N = 4n$ spins in Eq. 1. The crossing $E_σ = E_m$ is between the lowest singlet excitation and lowest triplet state; $^1E_{TT}$ is the singlet excitation that evolves from triplets on each sublattice at $J_1 = 0$. GS degeneracy and linear extrapolation are the $q_G$ points in the insets to Fig. 3. The critical points in the thermodynamic limit are at $P2 = -1.24$ and $P3 = 0.44$.

Figure 4 also shows the points in Table 1 at which the $q_G = π/2$ plateau is reached and left. The same linear extrapolation is used. Level crossing and GS degeneracy are consistent and independent determinations. Level crossing involves excited states while $q_G$ depends on GS correlations. Although P2 = –1.24(3) and P3 = – 0.44(2) are approximate, they indicate two IC phases in Fig. 1 separated by a C phase with $q_G = \pi/2$ and nondegenerate GS between $J_1/J_2$ = –1.24 and 0.44. The size dependence of level crossings at small $J_1$ is far weaker than that of $q_G$, but not as weak as at small $J_2$.

On the other hand, field theories that limit QLRO($\pi/2$) to $J_1 = 0$ would require all lines in Fig. 4 to extrapolate to $J_1 = 0$. This seems unlikely to us and in any case level crossing has to be taken into account. For whatever reason, field theories[12-14,18,23,28] that routinely refer to $E_\circ = E_m$ for P4 have not considered level crossing at small $J_1/J_2$. The present analysis of GS degeneracies is new evidence for a C phase with nondegenerate GS and $q_G = \pm \pi/2$ between two IC phases with doubly degenerate GS and variable $q_G$.

5. **Triplet state spin densities**

In this Section we further characterize the decoupled C phase using the triplet state $|T,\sigma\rangle$ with $S = S^z = 1$, excitation energy $E_m$ and inversion symmetry $\sigma$. ED gives $|T,\sigma\rangle$ explicitly in finite systems. The spin density at site r is

$$\rho_r(\sigma) = \langle T,\sigma|s_r^z|T,\sigma\rangle \tag{10}$$

The triplet's wave vector $k_T$ can be inferred from spin densities. Uniform $\rho_r = (4n)^{-1}$ at all sites indicates nondegenerate $|T,\sigma\rangle$ with $k_T = 0$ for $\sigma = 1$ and $\pi$ for $\sigma = -1$. Doubly degenerate $|T,\sigma\rangle$ with $\pm k_T$ in Eq. 2 indicates broken spin density symmetry with $\rho_r(1) \neq \rho_r(-1)$. The spin densities for $\sigma = \pm 1$ are proportional to $\cos^2 k_T r$ and $\sin^2 k_T r$, so that the sum $(4n)^{-1}$ is the same at all sites. The spin density of odd numbered sites is

$$\rho_A = \sum_{r=1}^{2n} \rho_{2r-1} \tag{11}$$

The sublattice spin density of even numbered sites is $\rho_B = 1 - \rho_A$ since $S^z = 1$ is conserved. We have $\rho_A = \rho_B = 1/2$ except when $k_T = \pm \pi/2$. At $J_1 = 0$, for example, $|T,\sigma\rangle$ is either $|T\rangle|G\rangle$ or $|G\rangle|T\rangle$ where $|G\rangle$ and $|T\rangle$ are the GS and lowest triplet of sublattices. The product functions have opposite $\sigma$ symmetry; $\rho_A$ or $\rho_B$ is uniformly $(2n)^{-1}$ on one sublattice and 0 on the other.

The evolution of $\rho_A \geq \rho_B$ is shown in Fig. 5 as a function of $J_1/J_2$ using ED up to $N = 28$. The sudden changes to $\rho_A = \rho_B = 1/2$ at $(J_1/J_2)_T$ indicate a level crossing of triplets. There are four degenerate triplets at $(J_1/J_2)_T$, a pair $|T,\pm 1\rangle$ with $k_T = \pi/2$ and a pair with $k_T = \pi/2 - \pi/2n$ in the $J_1 < 0$ sector or $\pi/2 + \pi/2n$ in the $J_1 > 0$ sector. Increasing $J_1/J_2 > 0$ or decreasing $J_1/J_2 < 0$ generates additional triplet degeneracies at which the sublattice spin densities do not change. The points $(J_1/J_2)_T$ are close to $(J_1/J_2)_n$ and $(J_1/J_2)_{n+1}$ where $q_G$ reaches and leaves the $\pi/2$ plateau. The IC phases have degenerate $|T,\sigma\rangle$ with $k_T < \pi/2$ or $> \pi/2$, while $|T,\sigma\rangle$ is nondegenerate in the dimer phase with $E_m > 0$ or in the QLRO($\pi$) phase with $E_m = 0$.

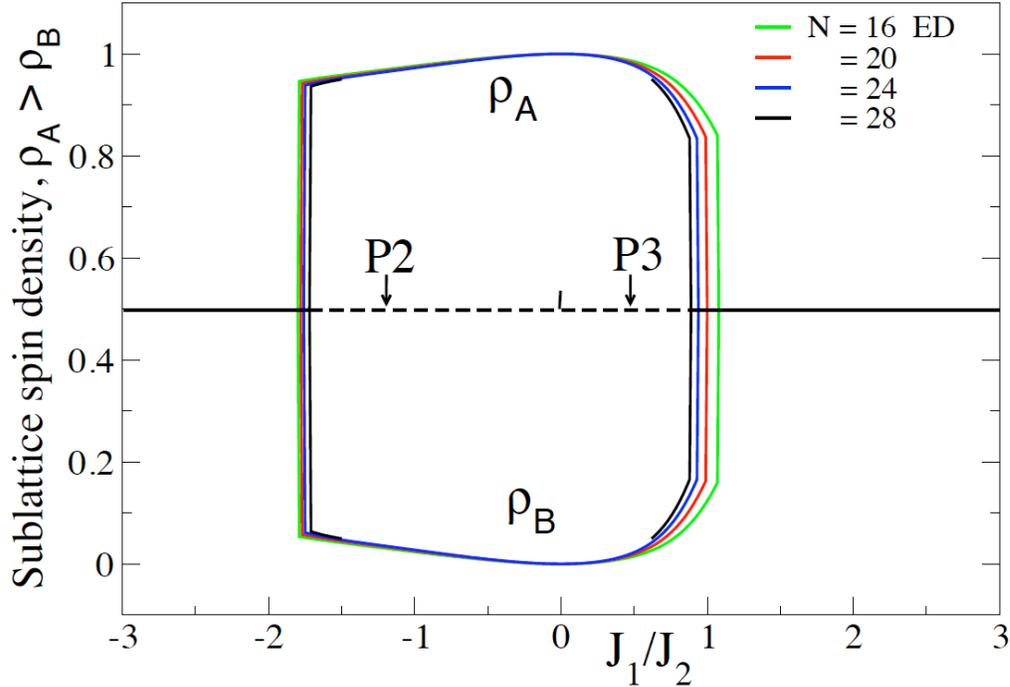

Fig. 5. Sublattice spin density, Eq. 11, with $\rho_A \geq \rho_B$ as a function of $J_1/J_2$ for N spins in Eq. 1. ED results to N = 28 show finite $|\rho_A - \rho_B|$ when the lowest triplet state $|T,\sigma\rangle$ has wave vector $k_T = \pm \pi/2$ and $\rho_A = \rho_B = 1/2$ otherwise. The critical points P2 and P3 are based on $q_G$ and $|\rho_A - \rho_B| = 1$ at $J_1 = 0$ is exact.

The eigenstates at $J_1 = 0$ are products of 2n-spin HAFs. The product basis is complete, and $|T,\sigma\rangle$ can expanded at any $J_1/J_2$ as a linear combination of products of sublattice eigenstates with $S_A + S_B = 1$. Since σ does not interchange sublattices, product functions have fixed σ. Since the sublattices are equivalent, however, the expansion coefficients $C_{ij}$ and $C_{ji}$ of products such as $^3|j\rangle^1|i\rangle$ and $^1|i\rangle^3|j\rangle$ must have equal magnitudes, where $|i\rangle$ and $|j\rangle$ refer to the ith singlet and jth triplet of 2n-spin HAFs. Triplets based on $S_A = S_B \geq 1$ have $\rho_A = \rho_B = 1/2$, while triplets based on $|S_A - S_B| = 1$ have $\rho_A \neq \rho_B$. Degenerate triplets $|T,\sigma\rangle$ at finite $J_1/J_2$ are linear combinations of product functions with decreasing but finite $|\rho_A - \rho_B|$ in Fig. 5 as long as $k_T = \pm \pi/2$.

DMRG is applicable to larger systems because $|T,\sigma\rangle$ has the lowest energy in the sector $S = S^z = 1$. The algorithm does not specify inversion symmetry, however, and returns some linear combination of $|T,1\rangle$ and $|T,-1\rangle$ for a degenerate triplet. There are four degenerate triplets at $(J_1/J_2)_T$ where sublattice spins become unequal. When $k_T = \pi/2$, we have $\rho_A \neq \rho_B$ *except* for the plus or minus linear combination. On the other hand, *all* linear combinations of $|T,1\rangle$ and $|T,-1\rangle$ lead to $\rho_A = \rho_B = 1/2$ when $k_T \neq \pi/2$. The wave vector is sufficient to estimate $(J_1/J_2)_T$ in large systems. Within numerical accuracy, finite $J_1/J_2$ leads to equal sublattice spin densities when $k_T \neq \pi/2$ and to finite fluctuations $|\rho_A - \rho_B| > 0$ when $k_T = \pi/2$. DMRG results for $(J_1/J_2)_T$ are consistent with but considerably less accurate than $(J_1/J_2)_n$ and $(J_1/J_2)_{n+1}$ in Table 1 for reaching and leaving the $q_G = \pi/2$ plateau. The $J_1 = 0$ spin densities for noninteracting sublattices are exact for any system size and $|\rho_A - \rho_B| > 0$ is readily demonstrated for $-1.2 < J_1/J_2 < 0.4$. The decoupled C phase identified by level crossing and GS degeneracy has a degenerate triplet $|T,\pm1\rangle$ with $k_T = \pm \pi/2$ and broken sublattice spin density $\rho_A \neq \rho_B$.

## 6. Magnitude of structure factor peaks

The spin structure factor, Eq. 5, has $C(0) = 3/4$ for $s = 1/2$ and satisfies the sum rule

$$\frac{1}{4n}\sum_q S(q) = \frac{3}{4} = \frac{1}{\pi}\int_0^\pi S(q)\,dq \tag{12}$$

The sum and integral refer to finite and infinite systems with discrete and continuous q. HAF spin correlations, indicated by the subscript zero, between distant sites go as[8,17]

$$C_0(r) \propto \frac{(-1)^r (\ln r)^{1/2}}{r} \tag{13}$$

Figure 6 shows $S(q)$ for HAFs and for the $J_1$-$J_2$ model at $J_1/J_2 = 2$. Since $(-1)^r C_0(r)$ is positive, the sum is over $|C_0(r)|$ and $S(\pi)$ diverges as $(\ln N)^{3/2}$ when the integral is cut off at N. The $S(\pi)$ peak increases as shown from $N = 24$ to 192. The size dependence is weak except at the peak, the area is conserved in all curves, and $S'(\pi)$ is not defined in the thermodynamic limit.

Only spins in the same sublattice contribute to $S(\pi/2)$. At $J_1 = 0$, the correlations between distant sites have even r in Eq. 13 and $\cos(r\pi/2)$ instead of $(-1)^r$. The GS has QLRO($\pi/2$) and divergent $S(\pi/2)$. The size dependence in Fig. 6 is again small except at the peaks $\pi/2$ and $3\pi/2$ ($= -\pi/2$). The $S(\pi/2)$ divergence is suppressed by $J_1$ in gapped IC phases. This is readily shown[29] for $1 < J_1/J_2 < 2$ where $E_m$ is large and correlations are short ranged.

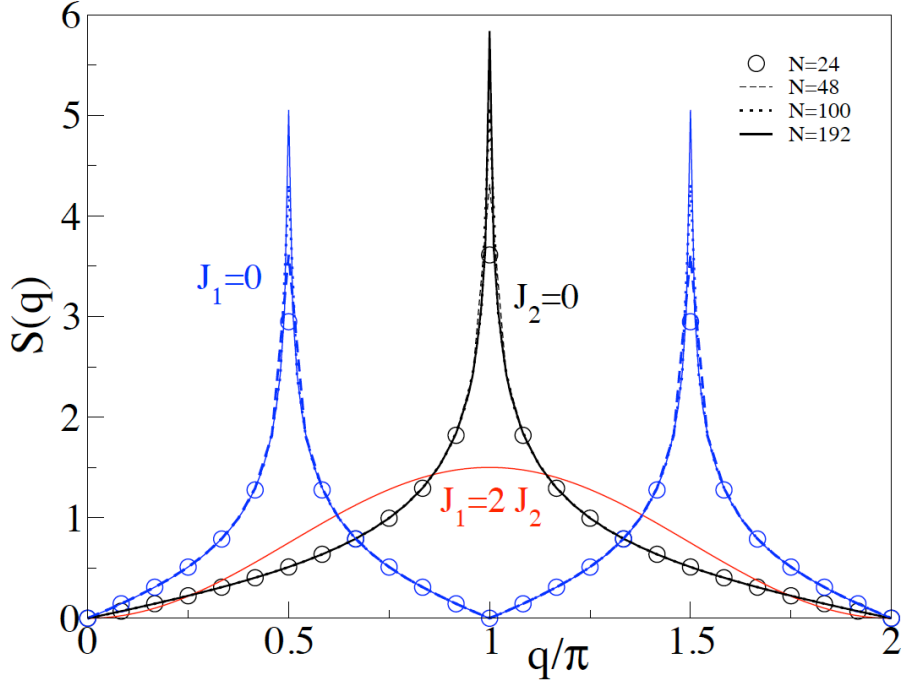

Fig. 6. Spin structure factor S(q), Eq. 4, for $J_2 = 0$, $J_1 = 0$ and $J_1/J_2 = 2$ in Eq. 1. Open symbols are ED for N = 24 and discrete wave vector q. Lines are DMRG for N = 48, 96 and 192, respectively, with increasing peaks and continuous q. S(q) at $J_1/J_2 = 2$ is Eq. 6, exact in the thermodynamic limit. The $\pi$ and $\pi/2$ peaks diverge in that limit.

Eq. 13 indicates divergent $S(\pi)$ at $J_2 = 0$ and $S(\pi/2)$ at $J_1 = 0$ in the thermodynamic limit. Divergent peaks are signatures of QLRO but have not been demonstrated directly for other parameters. Finite $S_{MG}(\pi) = 3/2$ clearly implies a critical point P4 such that $S(\pi)$ diverges for $J_1/J_2 \geq$ P4. The same logic leads to the critical points P2 and P3 where $S(\pi/2)$ diverges. We compute C(r) in finite systems and compare the size dependence of $S(\pi/2)$ at small $J_1$ to that of $S(\pi)$ at small $J_2$.

The magnitudes of $S(\pi; J_2 = 0, 4n)$ and $S(\pi/2; J_1 = 0, 8n)$ are necessarily equal since both systems are 4n-spin HAFs. Small $J_2$ gives a first-order correction, $J_2 C_0(2)$, that reduces $\pi$ order for $J_2 > 0$ and enhances it for $J_2 < 0$. Small $J_1$ couples noninteracting HAFs and there is no first-order correction. The difference between a perturbed system and weak exchange between two systems has important consequences. To lowest order in $J_1$, spin correlations within a sublattice go as[29]

$$C(2r) = C_0(2r) + O(J_1/J_2)^2 \tag{14}$$

The result holds in finite systems whether or not $C(2r)$ is amenable to exact evaluation. The leading terms in the Taylor expansion of $S(\pi;J_2,4n)$ about $J_2 = 0$ and of $S(\pi/2;J_1,8n)$ about $J_1 = 0$ are

$$\begin{aligned} S(\pi;J_2/J_1,4n) &= S(\pi;0,4n) - A_n(J_2/J_1) \\ S(\pi/2;J_1/J_2,8n) &= S(\pi;0,4n) - B_n(J_1/J_2)^2 \end{aligned} \tag{15}$$

ED for 24 spins returns[29] $A_6 \sim 10 B_3 > 0$ for $J_1/J_2 > 0$. DMRG to $N = 100$ confirms[29] that small $J_2/J_1$ reduces the $\pi$ peak more than small $J_1/J_2$ reduces the $\pi/2$ peak, which in turn is reduced faster for $J_1/J_2 > 0$ than for $J_1/J_2 < 0$. The initially quadratic dependence in Eq. 14 points to *weaker* suppression of the $S(\pi/2)$ divergence by $J_1$ than of the $S(\pi)$ divergence by $J_2$. Field theory asserts instead that $S(\pi/2)$ becomes finite for arbitrarily small $J_1/J_2$. Numerical analysis is consistent with divergent peaks in QLRO phases whose critical points are determined by level crossing and GS correlations.

There is a basic difference between the $\pi$ and $\pi/2$ peaks. $S(q)$ is symmetric about $\pi$, with $S(\pi - \varepsilon) = S(\pi + \varepsilon)$. If a finite system has a $\pi$ peak, the peak remains there when $S(q)$ is assumed to be continuous in the thermodynamic limit, whether or not the system size exceeds the range of spin correlations. In other words, whether or not the assumption is justified. The divergence is not related to the peak's position. The critical points P2 and P3 are between IC and QLRO($\pi/2$) phases, however, and $S(q)$ is not symmetric about $q = \pi/2$ except at $J_1 = 0$. A finite system with a $\pi/2$ peak has unequal $S(q)$ at $q = \pi/2 \pm \pi/2n$, as can readily be verified analytically. The peak necessarily shifts to $q < \pi/2$ for $J_1 < 0$ and continuous $S(q)$ or to $q > \pi/2$ for $J_1 > 0$ when $S(\pi/2)$ is *assumed* to be finite in the thermodynamic limit and hence differentiable at the peak. Even if accurate $C(r)$ could be found at small $J_1/J_2$ in large systems, extrapolation would be necessary to determine whether $S(\pi/2)$ diverges. Gapped IC phases have spin correlations of finite range, and $q_G$ immediately shifts at P2 or P3 from the $q_G = \pi/2$ plateau in Fig. 3.

**7. Sublattice spin and correlations**

The critical points P2 = –1.24 and P3 = 0.44 are far from symmetric about $J_1 = 0$. AF exchange between sublattices quickly induces an IC phase at P3 while the IC phase at P2 requires stronger F exchange. A qualitative explanation is that $J_1 > 0$ stabilizes the singlet $^1E_{TT}$ that is involved in level crossing and generates the GS degeneracy $(J_1/J_2)_{n+1}$ in systems of 4n spins. The eigenstate $^1|T\rangle|T\rangle$ at $J_1 = 0$ is the singlet linear combination of the lowest triplet on each sublattice. On the contrary, F exchange raises the energy of $^1|T\rangle|T\rangle$ and the singlet GS must be achieved with minimal sublattice spin. We present a more quantitative analysis of the P2, P3 asymmetry.

The GS of Eq. 1 with $J_2 \geq 0$ is a singlet for $-4 \leq J_1/J_2$. Sublattice spin is conserved at $J_1 = 0$ where $S_A = S_B = 0$, but not in general. The GS expectation value of $\langle S_A^2 \rangle = \langle S_B^2 \rangle$ per site is

$$\frac{\langle S_A^2 \rangle}{2n} \equiv \sum_{r=0}^{2n-1} C(2r) = \frac{S(\pi)}{2} = -2\sum_{r=1}^{n-1} C(2r-1) \tag{16}$$

The equality with $S(\pi)/2$ follows on using $S(0) = 0$ for singlets. The second equality is an immediate consequence of $\langle (S_A + S_B)^2 \rangle = 0$. The following results are exact: $\langle S_A^2 \rangle/2n = 0$ at $J_1 = 0$ as required; $\langle S_A^2 \rangle/2n$ is 3/4 at $J_1/J_2 = 2$ and it diverges for $J_1/J_2 \geq 4.148$. Another exact result obtained below is $\langle S_A^2 \rangle/2n = 1/4$ at $J_1/J_2 = -4$, three times smaller than at $J_1/J_2 = 2$. The size dependence is weak when $S(\pi)$ is a minimum. DMRG at $J_1/J_2 = -1$ and 0.4 returns $\langle S_A^2 \rangle/2n \sim 0.015$ and 0.010, respectively.

In units of $J_2$, the GS energy per site is $-3/4$ at $J_1/J_2 = -4$. The singlet correlations $C(r)$ go as $\cos(\pi r/2n)$ and satisfy two conditions: $C(0) = 3/4$ and $\Sigma_r C(r) = 0$. Up to amplitude $A(n)$, we have

$$C(r) = \frac{A(n)}{4}\cos(\pi r/2n) + \left(\frac{3}{4} - \frac{A(n)}{4}\right)\left(\frac{4n\delta_{0r} - 1}{4n - 1}\right) \tag{17}$$

where $\delta_{0r}$ is the Kroneker delta. The energy per site is $-4C(1) + C(2)$. Setting the energy equal to $-3/4$ gives a linear equation for $A(n)$. We obtain $A(n) = 1 + 1/(2n)$ up to corrections of order $n^{-4}$ and $\langle S_A^2 \rangle/2n = 1/4$. Adjacent spins are asymptotically parallel in the singlet. Eq. 17 is consistent with Hamada et al.[7] where a note added in proof indicates that an analytical expression for $\langle s_i^z s_j^z \rangle$ had been found at $J_1/J_2 = -4$.

We chose to study $J_1/J_2 = -1$ and 0.4 in the decoupled phase with $q_G = \pi/2$. Table 2 lists $C(r)$ up to $r = 20$ for $N = 96$ at $J_1/J_2 = -1, 0.4$ and 0, where $C_0(2r)$ refer to a 48-spin HAF. Shiroishi and Takahashi[39] obtained analytical expression for $C_0(r)$ in the thermodynamic limit up to $r = 4$. The first four entries $C_0(2r)$ at $N = 96$ differ from the analytical results by less than $10^{-3}$, and the $N = 192$ correlations by $< 2.5 \times 10^{-4}$. DMRG is quite accurate, as expected. Spin correlations $C(2r)$ within sublattices are almost identical at $J_1/J_2 = -1$ and 0.4. Their nodal structure goes as $\cos(\pi r)$, just as at $J_1 = 0$. Indeed, they remain close to $C_0(2r)$ as suggested by Eq. 14 even at substantial deviations from $J_1 = 0$. Exchange between sublattices leads to comparable but out of phase $C(2r-1)$ that follows from the fact that the GS energy has $J_1 C(1) < 0$.

Correlations are strictly limited to $r \leq 2n$ in systems of $N = 4n$ sites with PBC. Converged $C(r)$ are further limited to $r \sim n/2 = N/8$ based on various criteria.[17,40] DMRG results in Fig. 7 at $J_1/J_2 = -1$ and 0.4 show $r|C(r)|$ separately for even and odd $r$ up to $r = n$ for $4n = 64, 96$ and 144. Convergence is fair to $r < n/2$ along with typical HAF oscillations at small $r$, here on sublattices. Converged $r|C(2r)| \sim 0.25$ in Fig. 7 are roughly constant whereas the $r|C_0(2r)|$ at $J_1 = 0$ in Eq. 13 increases slowly.

Table 2. DMRG spin correlation functions $\langle s_0 \cdot s_r \rangle = C(r)$ to r = 20 for N = 96 spins in Eq. 1 and $J_1/J_2 = -1, 0.4$ and $0$.

| N = 96 | $J_1/J_2 = -1$ | $J_1/J_2 = 0.4$ | $J_1/J_2 = -1$ | $J_1/J_2 = 0.4$ | $J_1 = 0$ |
|---|---|---|---|---|---|
| r | C(2r-1) | C(2r-1) | C(2r) | C(2r) | $C_0(2r)$ |
| 1 | 0.02577 | -0.02141 | -0.43154 | -0.43771 | -0.44351 |
| 2 | -0.05097 | 0.02980 | 0.17342 | 0.17652 | 0.18238 |
| 3 | 0.03521 | -0.02220 | -0.13741 | -0.14199 | -0.15160 |
| 4 | -0.02960 | 0.01836 | 0.09029 | 0.09506 | 0.10487 |
| 5 | 0.02275 | -0.01498 | -0.07859 | -0.08305 | -0.09414 |
| 6 | -0.01934 | 0.01258 | 0.06035 | 0.06431 | 0.07490 |
| 7 | 0.01576 | -0.01045 | -0.05427 | -0.05801 | -0.06963 |
| 8 | -0.01331 | 0.00890 | 0.04498 | 0.04839 | 0.05917 |
| 9 | 0.01110 | -0.00744 | -0.04164 | -0.04488 | -0.05619 |
| 10 | -0.00928 | 0.00640 | 0.03663 | 0.03959 | 0.04963 |

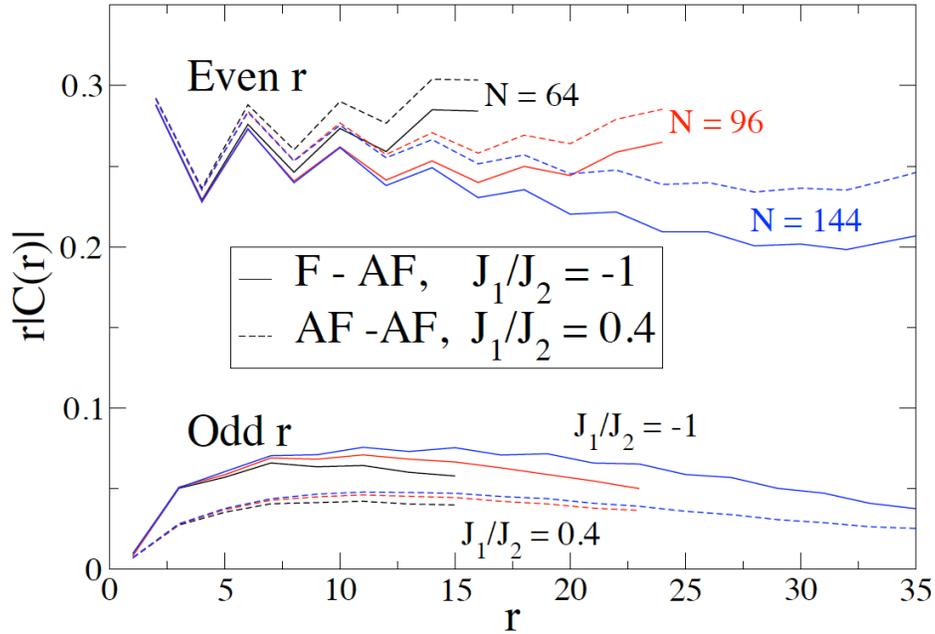

Fig. 7. Spin correlation functions C(r) in $J_1$-$J_2$ models with N spins in Eq. 1. Both $J_1/J_2 = -1$ and $0.4$ are in the decoupled phase with $q_G = \pi/2$ that includes $J_1 = 0$.

Substantial spin correlations $C(2r-1)$ are new results: at intermediate $r$ we find $(2r-1)|C(2r-1)| \sim 0.06$ and $\sim 0.04$ at $J_1/J_2 = -1$ and $0.4$, respectively. The nodal structure of $C(2r-1)$ in Table 2 goes as $\pm\sin((2r-1)\pi/2)$ with wave vector $\pi/2$. We understand $J_1 C(1) < 0$ and $q = \pi/2$ but not why the magnitude of $C(3)$ is larger than that of $C(1)$ or why $C(2r-1)$ then decreases roughly as $1/r$. The small value of $S(\pi)/2$ at $J_1/J_2 = -1$ or $0.4$ in the decoupled phase is due to extensive cancellation in Eq. 16 among spin correlations in different sublattices.

FM spin correlations $C(1) > 0$ are central to the "Haldane dimer" phase proposed by Furukawa et al.[28] in the interval $-4 \leq J_1/J_2 < 0$, as indicated in their Eq. 32 and shown in their Fig. 2 at $J_1/J_2 = -2$, $\Delta = 1$ (isotropic exchange), which is well inside the IC phase. The GS has slightly larger $C(1) > 0$ with one neighbor than the other. Such broken symmetry states can be constructed in finite $J_1$-$J_2$ models whenever the GS is doubly degenerate, and only inversion symmetry is broken in $J_1$-$J_2$ models with isotropic exchange.[27] The IC phases in Fig. 1 can also be viewed as dimer or bond-order-wave phases, both with $J_1 C(1) < 0$. Numerical results are shown in Fig. 6 of ref. 28 for $C(1)$, correlation lengths and string correlations. No points are shown, however, between $-1 < J_1/J_2 < 0.5$ which is considered to be an IC phase (except at $J_1 = 0$) that is beyond numerical analysis. The excluded region almost coincides with the QLRO($\pi/2$) phase between P2 and P3 in Fig. 1.

## 8. Two related models

In this Section we summarize two models whose quantum phases are related to those of the $J_1$-$J_2$ model. The first is an analytical model with HAFs on sublattices and mean-field exchange that suppresses IC phases and widens the QLRO($\pi/2$) phase; the critical points P1/P2 merge to $J_1/J_2 = -4\ln 2$ and P3/P4 to $J_1/J_2 = 4\ln 2$. The second retains $J_1$ between neighbors. Frustration within sublattices merges P2/P3 and generates a single

IC phase from $q_G = 0$ to $\pm \pi$. The motivation is to manipulate the critical points in Fig. 1 in predictable ways using exact thermodynamic results as far as possible.

The $J_1$-$J_2$ model has N exchanges $J_1$ between adjacent sites. The total exchange is the same for $(N/2)^2$ exchanges $4J_1/N$ between all spins in different sublattices, as in the Lieb-Mattis model.[41] Equal exchange $4J_1/N$ is the mean-field (mf) approximation for exchange between sublattices. The frustrated mf model for 4n spins is

$$H_{mf} = \frac{4J_1}{N} \sum_{r,r'=1}^{2n} \vec{s}_{2r} \cdot \vec{s}_{2r'-1} + J_2 \sum_{r=1}^{4n} \vec{s}_r \cdot \vec{s}_{r+2} \tag{18}$$

Sublattice spin is conserved as seen on rewriting the first term as

$$\frac{4J_1}{N} \sum_{r,r'=1}^{2n} \vec{s}_{2r} \cdot \vec{s}_{2r'-1} = \frac{2J_1}{N}\left((S_A + S_B)^2 - S_A^2 - S_B^2\right) \tag{19}$$

The eigenstates of $H_{mf}$ are products of HAF eigenstates in sectors with $S_A = S_B \le n$. We define $J_2 E(S,2n)$ as the lowest energy for $S \le n$. The $S = 0$ energy per site is $E(0,2n)/2n = \varepsilon_0 = 1/4 - \ln 2$ in the thermodynamic limit. The GS is the combination of $S = S_A + S_B$ and $S_A = S_B$ that minimizes the energy in Eq. 18.

The QLRO($\pi/2$) phase with $S = S_A = S_B = 0$ is the GS for $J_1 < 0$ until the FM state with $S = S_A + S_B = 2n$ is reached at $J_1/J_2 = -4\ln 2$ in the thermodynamic limit. The gapped IC phase with $J_1 < 0$ has disappeared and $q_G$ changes discontinuously from 0 to $\pi/2$ at $J_1/J_2 = -2.773$. The quantum transition is first order. The gapped IC and dimer phases are also suppressed for $J_1 > 0$. The AFM state with $S = 0$ and $S_A = S_B = n$ is reached in the thermodynamic limit at $J_1/J_2 = 4\ln 2$ where $q_G$ jumps from $\pi/2$ to $\pi$. In fact, the GS remains the product $|G\rangle|G\rangle$ of noninteracting HAFs between $J_1/J_2 = -4\ln 2$ and $\pi^2/4$; all $C(2r-1)$ are rigorously zero in the interval.[29] The GS between $\pi^2/4$ and $4\ln 2$ is[42] $^1|T\rangle|T\rangle$, the singlet linear combination of the lowest triplet of each sublattice.

There are infinitely many ways of going from equal $4J_1/N$ between spins in different sublattices and to nearest neighbor $J_1$. The critical points depend on the choice of exchange between sublattices. Large $J_1 < 0$ generates LRO(0) while large $J_1 > 0$ generates either[43,42] LRO($\pi$) or QLRO($\pi$). The mf model, Eq. 18, rigorously has a gapless critical QLRO($\pi/2$) phase. The critical points depend on the choice of exchanges, and *all* choices have QLRO($\pi/2$) at $J_1 = 0$ for decoupled sublattices. If somehow the QLRO($\pi/2$) phase of the $J_1$-$J_2$ model were limited $J_1 = 0$, the immediate question would be what exchange between sublattices restores the QLRO($\pi/2$) phase to a $J_1/J_2$ interval.

The second model is doubly frustrated. In addition to exchange $J_1$ between first neighbors, we consider exchange $J_4 > 0$ between second neighbors of sublattices. The doubly frustrated chain is

$$H_D = J_1 \sum_r \vec{s}_r \cdot \vec{s}_{r+1} + J_2 \sum_r \vec{s}_r \cdot \vec{s}_{r+2} + J_4 \sum_r \vec{s}_r \cdot \vec{s}_{r+4} \tag{20}$$

Several exact results follow immediately. When $J_1 = 0$, frustration is within sublattices and the MG point at $J_4/J_2 = 1/2$ is four-fold degenerate. The singlet-paired sites in one of the Kekulé diagrams of sublattice A is

$$|K1_A\rangle = (1,3)(5,7)\ldots(2n-3)(2n-1) \tag{21}$$

The other diagram goes as $(3,5)(7,9)\ldots(2n-1,1)$. The corresponding pairing in sublattice B is between nearest neighbor even sites. Still at $J_1 = 0$, the QLRO($\pi/2$) phase of either sublattice extends to $J_4/J_2 = 0.2411$, the critical point P4 for sublattices.

At constant $J_4/J_2 = 0.2411$, the quantum phase diagram of $H_D$ as a function of $J_1/J_2$ has a QLRO($\pi/2$) point at $J_1 = 0$ between two IC phases, and any additional frustration $J_1 \neq 0$ suppresses long-range spin correlations. The $J_1$-$J_2$ model has $J_4 = 0$ and its QLRO($\pi/2$) phase in Fig. 1 is suppressed at finite $J_1/J_2$. Let's consider the phase boundaries in the $J_1/J_2$, $J_4/J_2$ plane. The QLRO($\pi/2$) phase at the origin is stable along $J_1 = 0$ up to $J_4/J_2 = 0.2411$ and, in our analysis, to P2 < 0 and P3 > 0 when $J_4 = 0$. Field theories limit the QLRO($\pi/2$) phase of the $J_1$-$J_2$ model to $J_1 = 0$. The implied fragility of the QLRO($\pi/2$) phase to $J_1 \neq 0$ at $J_4 = 0$ is contrasted in the doubly frustrated model its robustness at $J_1 = 0$ to $J_4 > 0$.

The $J_4$ term of $H_D$ strongly perturbs the points at which the GS is degenerate. ED for N = 24 returns the $q_G$ vs. $J_1/J_2$ staircases in Fig. 8 for $J_4/J_2 = 0$, 1/4, and 1/2. The inset enlarges where $q_G = \pi/2$ is reached at $J_1 < 0$ and left at $J_1 > 0$. Increasing $J_4$ substantially lengthens the steps on the $J_1 < 0$ side and decreases the $\pi/2$ plateau, which at $J_4/J_2 = 1/4$ is no wider than some other steps. As shown in the inset, the $q_G = \pi/2$ step is almost entirely suppressed at $J_4/J_2 = 1/2$, the MG point of noninteracting sublattices. Finite $J_1$ breaks the four-fold degeneracy. The S(q) peak at $J_4/J_2 = 1/2$ remains at $\pi/2$ for $J_1/J_2 = \pm 10^{-5}$ but is already at $\pi/2 \pm \pi/12$ at $J_1/J_2 = \pm 10^{-4}$.

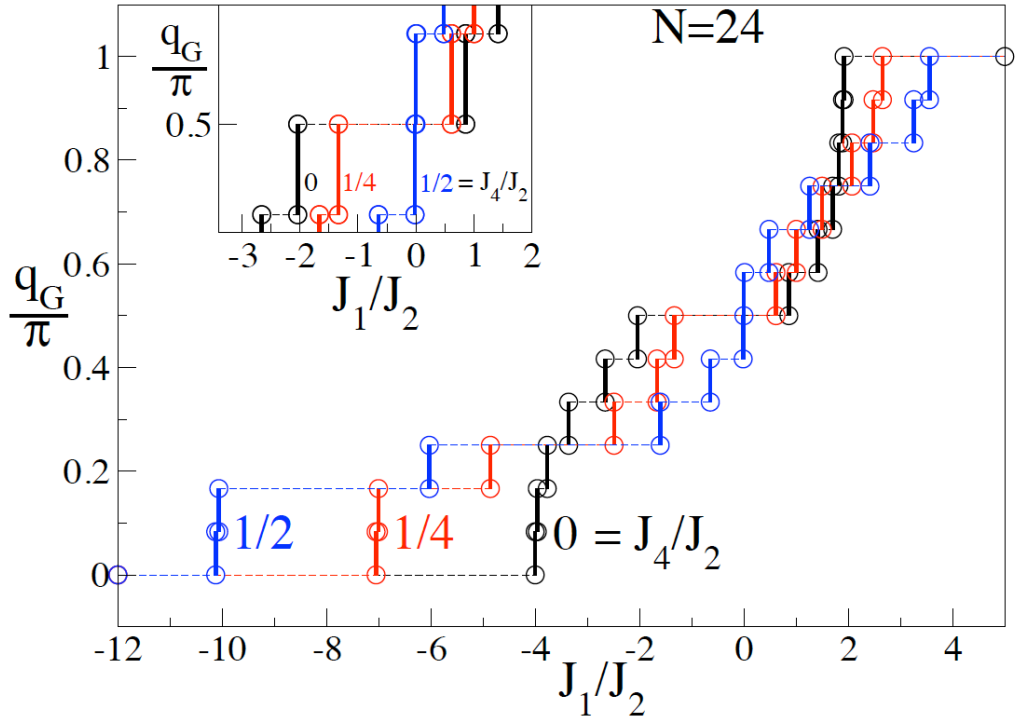

Fig. 8. ED results for the wave vector $q_G$ of GS correlations of the doubly frustrated model, Eq. 20, with N = 24 spins and $J_4/J_2$ = 0, 1/4 and 1/2. The inset enlarges the $q_G$ = $\pi/2$ region. The singlet GS is degenerate at 12 points in sectors that are even and odd under inversion at sites.

Increasing $J_4$ shifts the GS degeneracy with the ferromagnetic state to more negative $J_1/J_2$. Classical spins provide a qualitative explanation. The extra term $J_4\cos4q_{cl}$ leads to pitch angle $q_{cl}$

$$\frac{J_1}{J_2} = -4\left(1+\frac{4J_4}{J_2}\cos2q_{cl}\right)\cos q_{cl} \tag{22}$$

The $q_{cl}$ = 0 result is $J_1/J_2$ = − 4, exact at $J_4$ = 0. It increases to $J_1/J_2$ = − 8 and −12 at $J_4/J_2$ = 1/4 and 1/2, which is more negative than for quantum spins. The slope $(\partial q_{cl}/\partial(J_1/J_2))_0$ at $J_1$ = 0 is 1/4 when $J_4$ = 0. The slope diverges at $J_4/J_2$ = 1/4, when the r.h.s. of Eq. 22 is − $8\cos^3 q_{cl}$. For classical spins and $J_4/J_2$ = 1/2, $q_{cl}$ jumps discontinuously from $\pi/3$ to $2\pi/3$ at $J_1$ = 0, as follows from $4\cos^2 q_{cl}$ = 1. The N = 24 results for $q_G$ in Fig. 8 are consistent with the expectation that the QLRO($\pi/2$) phase of $H_D$ is suppressed in the thermodynamic limit for $J_4/J_2$ > 02411.

## 9. Discussion

The GS of the $J_1$-$J_2$ model, Eq. 1 with $J_2 \geq 0$, is a singlet for − 4 ≤ $J_1/J_2$. Other spin-1/2 chains with frustrated isotropic exchange have a singlet GS over some range of parameters. The singlet GS of finite systems with PBC is nondegenerate in general, but is doubly degenerate at 2n points in models with 4n spins and inversion symmetry at sites. The wave vector $q_G$ of spin correlations can be used to find GS degeneracies in IC phases, which is our principal result. Variable $q_G$ in Fig. 3 indicates two IC phases. One is between the exact critical point $J_1/J_2$ = − 4 and $J_1/J_2$ = −1.24 based on the size dependence of $q_G$; the other is between $J_1/J_2$ = 0.44 estimated from $q_G$ and the exact MG point, $J_1/J_2$ =

2, which is the C-IC point. In between is a gapless critical C phase with nondegenerate GS and QLRO($\pi/2$). The lowest triplets $|T,\pm1\rangle$ in the decoupled phase have wave vector $k_T = \pm \pi/2$ and broken sublattice spin densities $\rho_A \neq \rho_B$ that reaches $|\rho_A - \rho_B| = 1$ at $J_1 = 0$.

The structure factor $S(q)$ is a convenient way to find energy degeneracies in finite systems using GS properties. The GS is rigorously nondegenerate on the $q_G = \pi/2$ plateaus between $(J_1/J_2)_n$ and $(J_1/J_2)_{n+1}$ in Table 1. Our numerical results are in excellent agreement with previous results for $S(q)$ in other contexts. As noted in Section 3, the Lifshitz point where $S''(\pi) = 0$ is $(J_1/J_2)_L = 0.52066$, which matches the result of Bursill et al.[11] Sudan et al.[24] studied multipolar spin correlations and magnetization of the $J_1$-$J_2$ model with $J_1 < 0$. The lower panel of their Fig. 4 shows the $S(q)$ peak, $q_{max}/\pi$, for $-4 \leq J_1/J_2 \leq -2$ at zero field based on ED to N = 28. Our DMRG results leading to Eq. 8 are closely similar: $q_{max}/\pi = 0.250$ at $J_1/J_2 = -3.53$ and $\sim 0.45$ at $-2$ in either case. Furukawa et al.[28] use the infinite time evolving block decimation algorithm (iTEBD). The $\Delta = 1$ (isotropic exchange) curve of the $S(Q)$ peak Q vs. $J_1/J_2$ in Fig. 16 of ref. 28 has constant Q = $\pi/2$ in the $J_1/J_2$ interval from about $-1.1$ to $0.5$. The resemblance to $q_G/\pi$ in Fig. 3 is striking. In the IC phase at $J_1/J_2 = -1.8$, Fig. 8 of ref. 28, has $Q/\pi = 0.470$ compared to 0.466 according to Eq. 8.

We summarize the quantum phase diagram of the $J_1$-$J_2$ model in Fig. 1 as follows. The gapless FM phase with LRO(0) holds in the sector with $J_1 < 0$ and $J_2/J_1 \leq -1/4$, including $J_2 < 0$. Similarly, the gapless AFM phase with QLRO($\pi$) holds in the sector $J_1 > 0$ and $J_2/J_1 \leq 0.2411$. The gapless decoupled phase with QLRO($\pi/2$) holds in the sector with $J_2 > 0$ and $-1.24 \leq J_1/J_2 \leq 0.44$. Between the gapless phases with nondegenerate GS are gapped IC and dimer phases with doubly degenerate GS and spin correlations of finite range. The IC phase with $-4 \leq J_1/J_2 \leq -1.24$ has variable $q_G$ ranging from 0 to $\pm \pi/2$. The IC phase with $0.44 \leq J_1/J_2 \leq 2$ has $q_G$ ranging from $\pm \pi/2$ to $\pi$ (= $-\pi$). The dimer phase has $q_G = \pi$, and $2 \leq J_1/J_2 \leq 4.148$. As seen in Fig. 4, level crossing and GS degeneracy extrapolate to the same critical points P2 = $-1.24$ and P3 = 0.44. The gapless QLRO(q) phases have divergent $S(q)$ peaks while gapped phases have finite peaks.

These numerical results have a simple qualitative interpretation. The GS energy per site of the $J_1$-$J_2$ model is

$$\varepsilon_o(J_1, J_2) = J_1 C(1) + J_2 C(2) \tag{23}$$

The GS at $J_2 = 0$ has spin correlations with $q_G = 0$ for $J_1 < 0$ and $q_G = \pi$ for $J_1 > 0$. The second neighbor correlation $C(2)$ is positive in both cases. It follows that $C(1)$ changes sign when $J_1$ does and that $J_2 > 0$ is frustrating in either case. Increasing $J_2$ leads to $C(2) = 0$ at the MG point, $J_1/J_2 = 2$, or at $q_G = \pi/4$ where $J_1/J_2 = -3.53$. The situation is quite different in the decoupled QLRO($\pi/2$) phase in which spin correlations $C(2r–1)$ between sublattices are identically zero at $J_1 = 0$. Since the phase is compatible with small $C(1)$ of either sign, finite $J_1$ is very weakly frustrating at first and the decoupled phase extends over a substantial interval about $J_1 = 0$.

As proposed by White and Affleck,[12] the distance dependence of spin correlations in gapped IC phases goes as

$$C_{IC}(r) \propto (\cos Qr) r^{-1/2} \exp(-r/\xi) \tag{24}$$

They remark that Eq. 24 is approximate and holds for $r/\xi \gg 1$. DMRG calculations of $C_{IC}(r)$ at $J_1/J_2 = 0.56$, well inside the IC phase, were fit[12] in their Fig. 9 to $\xi = 17.1$ and pitch angle $Q = \pi/2 + \pi/(4\xi)$. DMRG with two spins added per site has numerical difficulties[12] when $J_1/J_2 < 0.5$. The field theory of White and Affleck leads[12] to $1/\xi \propto \exp(-aJ_2/J_1)$ for $J_1 > 0$, where a is a free parameter, while that of Itoi and Qin returns[14] $1/\xi \propto \exp(-c(J_2/J_1)^{2/3})$ with different c for positive and negative $J_1$. In either case, $\xi > 0$ ensures a finite range of correlations and hence a finite $S(q)$ peak for $J_1 \neq 0$.

The expression for $C_{IC}(r)$ has been adopted and rationalized in subsequent studies of the $J_1$-$J_2$ model[33,28] as well as the bilinear-biquadratic chain of S = 1 spins.[30] Now Q is identified as q*, the structure factor peak. Furukawa et al.[28] report (Fig. 6b of ref. 28) $\xi$ = 36 at $J_1/J_2$ = −1.8 in the IC phase; they note that, as anticipated by Itoi and Qin,[14] $\xi$ is larger for $J_1 < 0$ than for $J_1 > 0$. Still in the IC phase, DMRG with four spins added[16] per step leads to $\xi$ = 27 and 23.5 at $J_1/J_2$ = 0.48 and 0.54, respectively, and as shown in Fig. 6 of ref. 16 requires different amplitudes for C(2r) within and C(2r–1) between sublattices. These examples indicate that $C_{IC}(r)$ holds in IC phases. However, the C(r) in Table 2 are not compatible with $C_{IC}(r)$ and Q = $\pi/2$, which immediately gives C(2r–1) = 0, in contrast to finite correlations in Fig. 8 between spins in different sublattices and $q_G$ = $\pi/2$ at both $J_1/J_2$ = −1 and 0.4. These $J_1/J_2$ parameters are in the decoupled C phase with QLRO($\pi/2$) rather than in a gapped IC phase.

We have presented numerical evidence for the quantum critical points P2 and P3 in Fig. 1 between gapped IC phases and a gapless decouple phase with QLRO($\pi/2$). First, GS spin correlations yield the structure factor S(q) whose peak $q_G$ tracks energy degeneracy. The GS is nondegenerate with $q_G$ = $\pi/2$ between P2 = $J_1/J_2$ = −1.24 and P3 = $J_1/J_2$ = 0.44 as shown in Table 1 and Fig. 3 using DMRG up to N = 192 spins. Second, level crossing discriminates between systems whose lowest excitation is a singlet or triplet. Exact level crossings in Fig. 4 up to N = 28 spins yield the same critical points as $q_G$. Third, the lowest triplet has $k_T$ = $\pi/2$ and sublattices spin densities $\rho_A \neq \rho_B$ in the decoupled C phase. Fourth, gapless critical phases have divergent S(q) peaks, with q = $\pi$ for $J_1/J_2 \geq$ P4 = 4.148 in the familiar QLRO($\pi$) phase and q = $\pi/2$ in the C phase between P2 and P3. Extrapolation to the thermodynamic limit is also required for the divergence of S($\pi$) or S($\pi/2$). The related models in Section 8 are additional evidence for a decoupled phase with QLRO($\pi/2$) in spin-1/2 chains with noninteracting sublattices at $J_1$ = 0.

We mentioned in the Introduction that weak exchange $J_1$ between quantum systems presents challenges with some resemblance to dispersion forces that, for example, have been difficult to include in density functional theory. Methods that are suitable at small frustration $J_2/J_1$ may be less effective at small $J_1/J_2$. Field theories extend a finite energy gap to $J_1$ = 0 on the basis of RG flows. The continuum limit of the lattice is an approximation and there are other approximations as well. Field theory has not so far addressed level crossing of excited states, variable $q_G$ of GS spin correlations in IC phases, spin densities of the lowest triplet or the magnitude of $S(q)$ peaks in connection with the critical points P2 and P3. We anticipate that the field theory of $J_1$-$J_2$ model will eventually be as consistent with numerical results in the sector of small $J_1/J_2$ as it is for the critical point P4 and dimer phase at small $J_2/J_1$.

**Acknowledgements** : We thank D. Sen, A.W. Sandvik and S. Ramasesha for instructive discussions and the NSF for partial support of this work through the Princeton MRSEC (DMR-0819860). MK thanks DST for a Ramanujan Fellowship and support for thematic unit of excellence on computational material science.